\documentclass[preprint,preprintnumbers,amsmath,amssymb, showpacs]{revtex4}
\usepackage{latexsym,makeidx}
\usepackage{epsfig, color, ulem}
\usepackage{setspace}

\def\rev#1{\textcolor{black}{#1}}

\begin{document}

\title{An acoustic imaging method for layered non-reciprocal media}

\author{Kees Wapenaar and Christian Reinicke}
\affiliation{Department of Geoscience and Engineering, Delft University of Technology, Stevinweg 1, 2628 CN Delft, The Netherlands}

\date{\today}

\begin{spacing}{1.3}

\begin{abstract}
Given the increasing interest for non-reciprocal materials, we propose a novel acoustic imaging method for layered non-reciprocal media. 
The method we propose is a modification of the Marchenko imaging method, which handles multiple scattering between the layer interfaces in a data-driven way. 
We start by reviewing the basic equations for wave propagation in a non-reciprocal medium. 
Next, we discuss Green's functions, focusing functions, and their mutual relations, for a non-reciprocal horizontally layered medium. 
These relations form the basis for deriving the modified Marchenko method, 
which retrieves the wave field inside the non-reciprocal medium from reflection measurements at the boundary of the medium. 
With a numerical example we show that the proposed method is capable of imaging the layer interfaces at their correct positions, 
without artefacts caused by multiple scattering.    
\end{abstract}


\maketitle

\bibliographystyle{eplbib}
\def\a{\rm (c)}
\def\ssss{s}
\def\sigmaa{p}
\def\bN{{\bf N}}
\def\bJ{{\bf J}}
\def\barrho{\rho^o\!\!\!}
\def\ba{{{\mbox{\boldmath ${\cal A}$}}}}
\def\bA{\,\,\,\tilde{\!\!\!\ba}}
\def\bc{{{\mbox{\boldmath ${\cal B}$}}}}
\def\bB{\,\tilde{\!\bc}}
\def\bl{{{\mbox{\boldmath ${\cal L}$}}}}
\def\bL{\,\tilde{\!\bl}}
\def\bh{{{\mbox{\boldmath ${\cal H}$}}}}
\def\bH{\,\tilde{\!\bh}}
\def\bq{\tilde{\bf q}}
\def\bp{\tilde{\bf p}}
\def\u{\tilde u}
\def\uu{u}
\def\m{m}
\def\s{\vartheta}
\def\ss{d_\alpha}
\def\bb{b}
\def\h{\s_{33}}
\def\p{\ssss_\alpha}
\def\q{\ssss_3}
\def\bx{{\bf x}}
\def\i{i}

\section{Introduction}

Currently there is an increasing interest for elastic wave propagation in non-reciprocal  materials \cite{Willis2012CRM, Norris2012RS, Trainiti2016NJP, Nassar2017JMPS, Attarzadeh2018JSV}.
We propose a novel method that uses the single-sided reflection response of a layered non-reciprocal medium to form an image of its interior. 
Imaging of layered media is impeded by multiple scattering between the layer interfaces. Recent work, building on the Marchenko equation \cite{Marchenko55DAN}, has led to
imaging methods that account for multiple scattering in 2D and 3D inhomogeneous media \cite{Broggini2012EJP, Wapenaar2013PRL, Neut2016GEO, Ravasi2016GJI}. 
Here we modify Marchenko imaging for non-reciprocal media. 
We restrict ourselves to horizontally layered media, but the proposed method can be generalised to 2D and 3D inhomogeneous media 
in a similar way as has been done for reciprocal media in the aforementioned references.

\section{Wave equation for a non-reciprocal medium}

For simplicity, in this paper we approximate elastic wave propagation  by an acoustic wave equation. 
Hence, we only consider compressional waves and ignore  the conversion from compressional waves to shear waves and vice versa. 
This approximation is often used in reflection imaging methods and is acceptable as long as the propagation angles are moderate.

\begin{figure}
\centerline{\epsfysize=7 cm \epsfbox{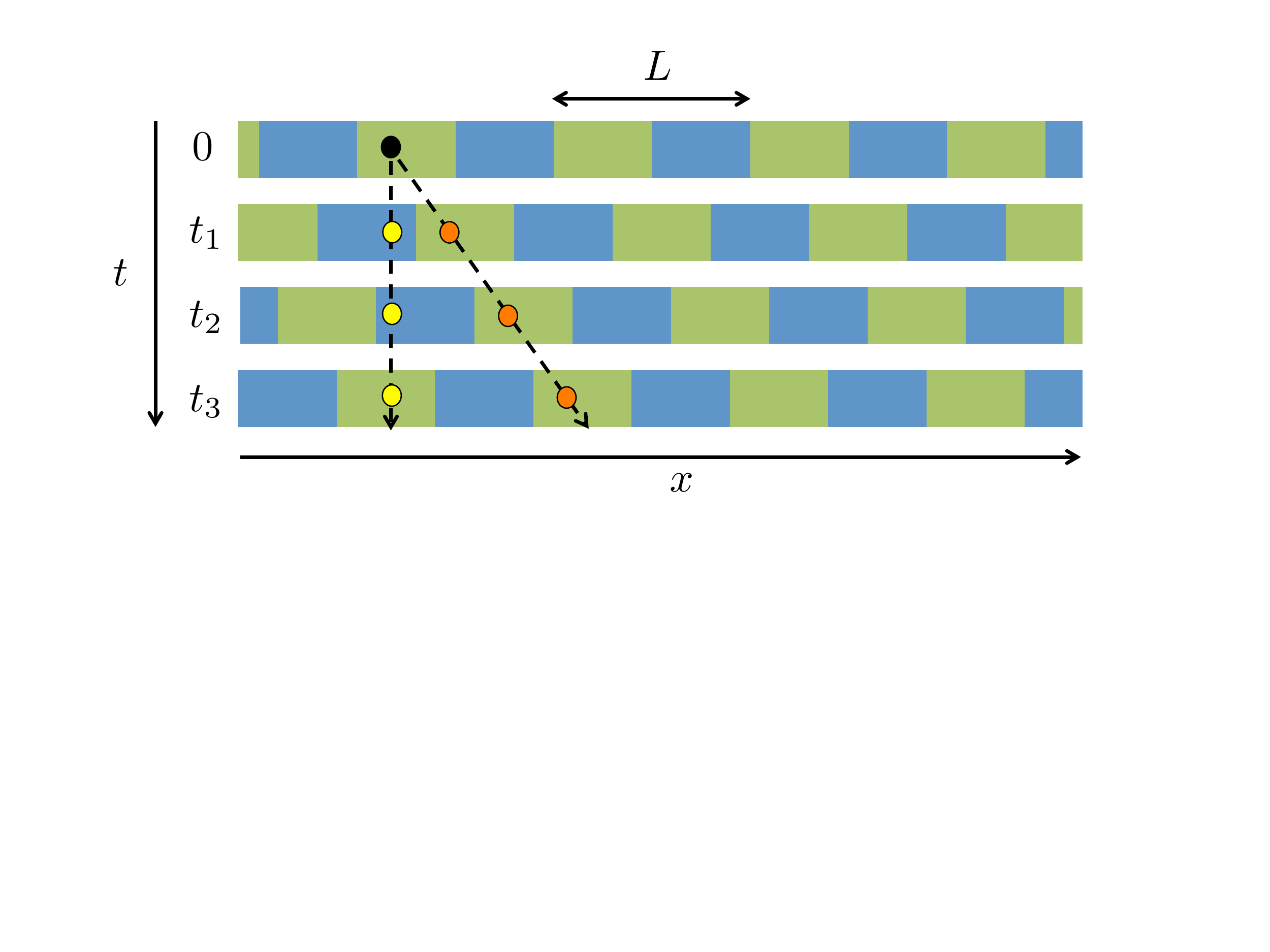}}
\vspace{-3.2cm}
\caption{\footnotesize A modulated 1D phononic crystal (after Nassar et al.  \cite{Nassar2017JMPS}). 
An observer at a fixed spatial position, indicated by the yellow dots, experiences a time-dependent medium, whereas an observer moving along with 
the modulating wave, indicated by the red dots, experiences a time-independent medium.}
\label{Figure1}
\end{figure}

We review the basics of non-reciprocal acoustic wave propagation. For a more thorough discussion we refer to the citations given in the introduction.
An example of a non-reciprocal material is a phononic crystal of which the parameters are modulated in a wave-like fashion \cite{Nassar2017JMPS}.
Figure \ref{Figure1} shows a modulated 1D phononic crystal at a number of time instances. The different colours represent different values 
of a particular medium parameter, for example the compressibility $\kappa$. 
This parameter varies as a function of space and time, according to $\kappa(x,t)=\kappa(x-c_m t)$, where $c_m$ is the modulation speed. 
The modulation wavelength is $L$. We define a moving coordinate $x'=x-c_m t$. The parameter $\kappa$ in the moving coordinate system, $\kappa(x')$,  is a function of space only.
The same holds for the mass density $\rho(x')$. 
Acoustic wave propagation in a modulated material is analysed in a moving coordinate system, hence, in a time-independent medium.
In this paper we assume the modulation speed is smaller than the lowest acoustic wave propagation velocity. Moreover, for the acoustic field we consider low frequencies, 
so that the wavelength of the acoustic wave is much larger than the modulation wavelength $L$. Using homogenisation theory, the small-scale parameters of the modulated material can be replaced
by effective medium parameters. The theory for 3D elastic wave propagation in modulated materials, including the homogenisation procedure, 
is extensively discussed by Nassar et al.  \cite{Nassar2017JMPS}.
Here we present the main equations (some details are given in the supplementary material). 
We consider a coordinate system $\bx=(x_1,x_2,x_3)$ that moves along with
the modulating wave (for notational convenience we dropped the primes). The $x_3$-axis is pointing downward.
In this moving coordinate system the macroscopic acoustic deformation equation and equation of motion for a lossless non-reciprocal material
read (leading order terms only)
\begin{eqnarray}
 \kappa\partial_t \sigmaa +(\partial_i+\xi_i\partial_t)v_i&=&0,\label{eqbb99}\\
(\partial_j+\xi_j\partial_t) \sigmaa+ {\barrho}_{jk}\partial_tv_k&=&0.\label{eqbb98}
\end{eqnarray}
Operator $\partial_t$ stands for temporal differentiation and $\partial_i$ for differentiation in the $x_i$-direction. Latin subscripts (except $t$) take on the values 1 to 3.
Einstein's summation convention applies to repeated Latin subscripts, except for subscript $t$. Field quantities $p=p(\bx,t)$ and $v_i=v_i(\bx,t)$ are the macroscopic
acoustic pressure and particle velocity, respectively. 
Medium parameters $\kappa=\kappa(\bx)$ and ${\barrho}_{jk}={\barrho}_{jk}(\bx)$ are the effective compressibility and mass density, respectively.
Note that the effective mass density may be anisotropic, even when it is isotropic at the micro scale. It obeys the symmetry relation $\barrho_{jk}=\barrho_{kj}$. Parameter $\xi_i=\xi_i(\bx)$ is an effective coupling parameter.

We obtain the wave equation for the acoustic pressure $\sigmaa$ by eliminating the particle velocity $v_i$ from equations (\ref{eqbb99}) and (\ref{eqbb98}). To this end,
define $\s_{ij}$ as the inverse of $\barrho_{jk}$, hence, $\s_{ij}\barrho_{jk}=\delta_{ik}$, where $\delta_{ik}$ is the Kronecker delta function.
Note that $\s_{ij}=\s_{ji}$.
Apply $\partial_t$ to equation (\ref{eqbb99}) and $(\partial_i+\xi_i\partial_t)\s_{ij}$ to
equation (\ref{eqbb98}) and subtract the results. This gives
\begin{eqnarray}
&& (\partial_i+\xi_i\partial_t) \s_{ij}(\partial_j+\xi_j\partial_t)\sigmaa-\kappa\partial_t^2\sigmaa=0.
\label{eq16aghht}
\end{eqnarray}
As an illustration, we consider a homogeneous isotropic effective medium, with $\s_{ij}=\delta_{ij}\rho^{-1}$. For this situation the wave equation simplifies to
\begin{eqnarray}
&& (\partial_i+\xi_i\partial_t)(\partial_i+\xi_i\partial_t)\sigmaa-\frac{1}{c^2}\partial_t^2\sigmaa=0,
\label{eq16aghhthom}
\end{eqnarray}
with $c=1/\sqrt{\rho\kappa}$. Consider a plane wave $p(\bx,t)=p(t-s_ix_i)$, with $s_i$ being the slowness in the $x_i$-direction. Substituting this into equation (\ref{eq16aghhthom}) we find the
following relation for the slowness surface
\begin{eqnarray}
(s_1-\xi_1)^2+(s_2-\xi_2)^2+(s_3-\xi_3)^2=\frac{1}{c^2},
\end{eqnarray}
which describes a sphere with radius $1/c$ and its centre at $(\xi_1,\xi_2,\xi_3)$. 
The asymmetry of this sphere with respect to the origin $(0,0,0)$ is a manifestation of the non-reciprocal properties  of the medium.

\section{Green's functions and focusing functions}

\begin{figure}
\vspace{-.cm}
\centerline{\epsfysize=9 cm \epsfbox{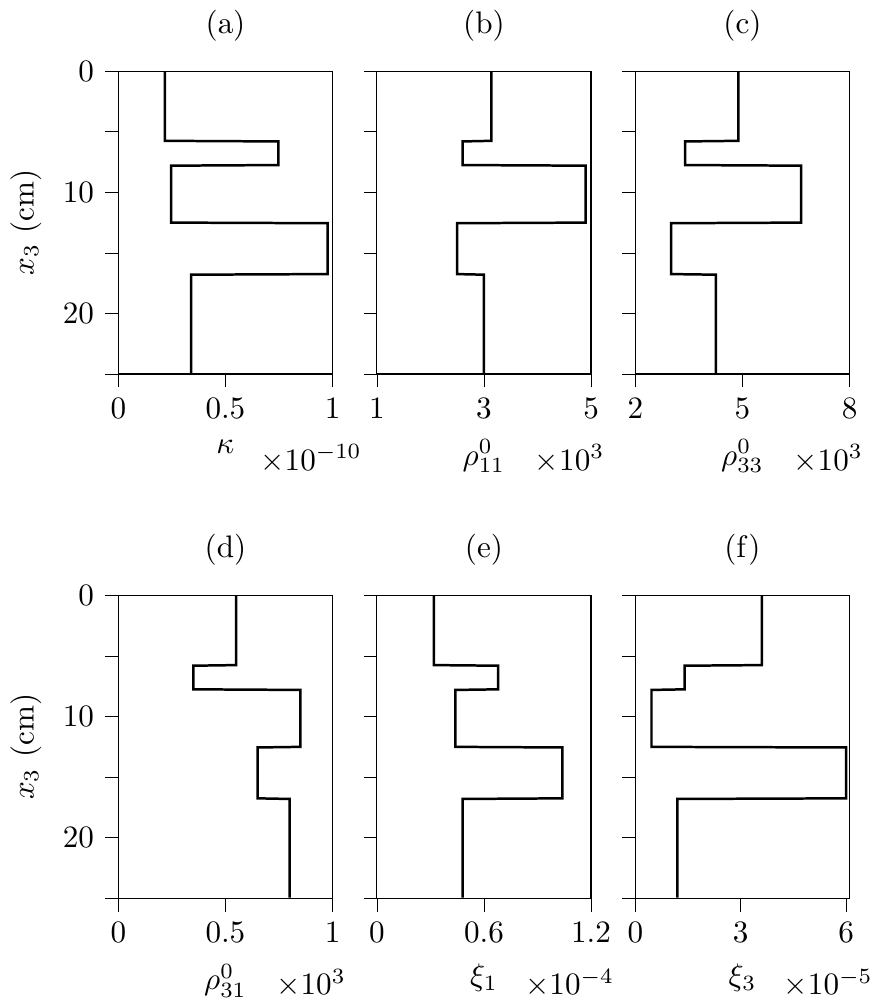}}
\vspace{-0.3cm}
\caption{\footnotesize Parameters of the non-reciprocal layered medium.}
\label{Fig3}
\end{figure}

The Marchenko method, which we discuss in the next section, makes use of specific relations between Green's functions and focusing functions.
Here we introduce these functions for a lossless non-reciprocal horizontally layered acoustic medium at the hand of a numerical example. 
Figure \ref{Fig3} shows the parameters of the layered medium as a function of the depth coordinate $x_3$. 
The half-space above the upper boundary $x_{3,0}=0$ is homogeneous.
For convenience we consider wave propagation in the $(x_1,x_3)$-plane (where $x_1$ and $x_3$ are moving coordinates, as discussed in the previous section).
Hence, from here onward subscripts $i$, $j$ and $k$ in equations  (\ref{eqbb99}) and (\ref{eqbb98}) take on the values 1 and 3 only.

\begin{figure}
\centerline{\epsfysize=15 cm \epsfbox{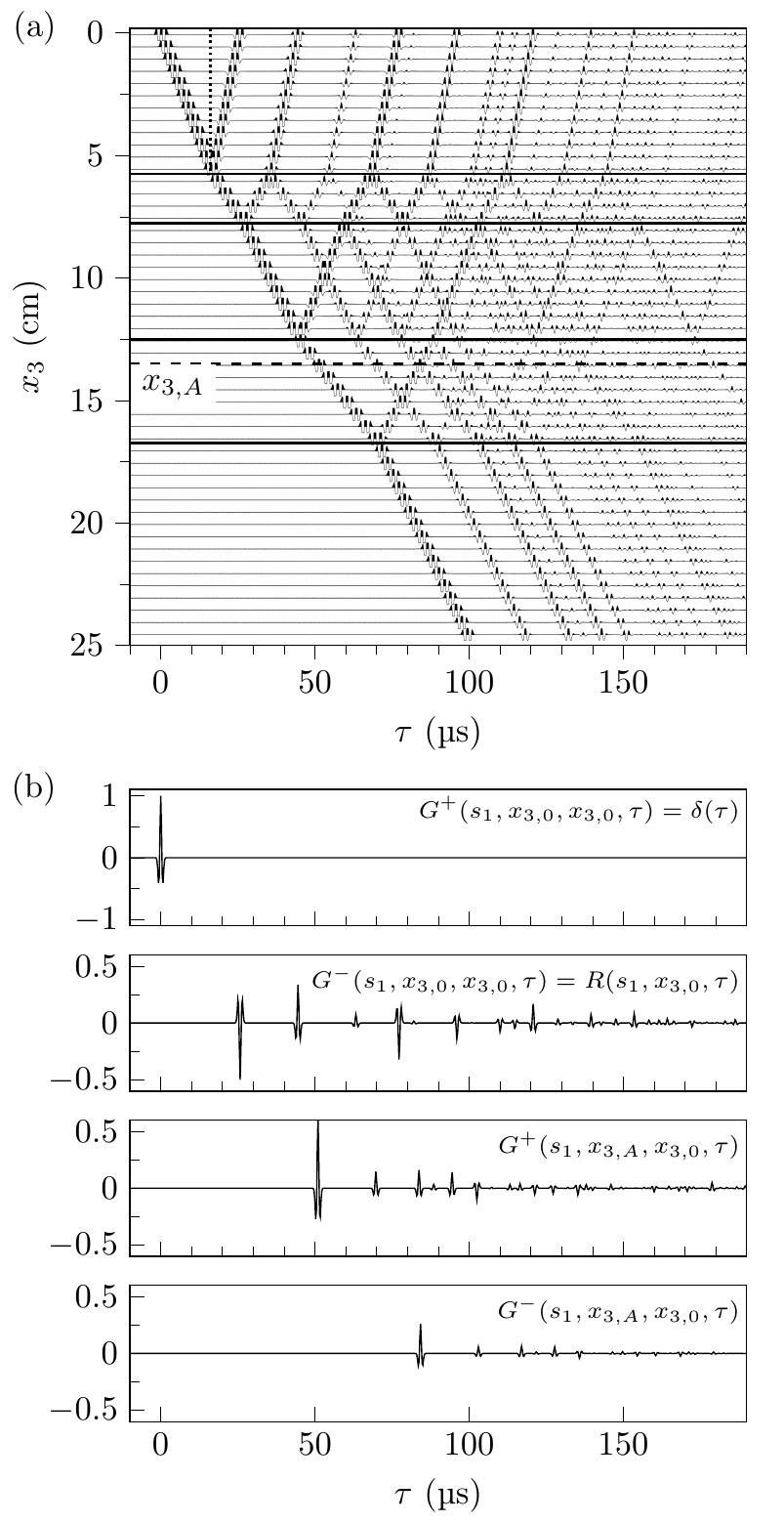}}
\caption{\footnotesize (a) Green's function $G(s_1,x_3,x_{3,0},\tau)$, for $s_1=0.22$ ms/m. (b) Decomposed Green's functions at $x_{3,0}=0$ and $x_{3,A}$.}
\label{Figure2}
\end{figure}

\begin{figure}
\centerline{\epsfysize=15 cm \epsfbox{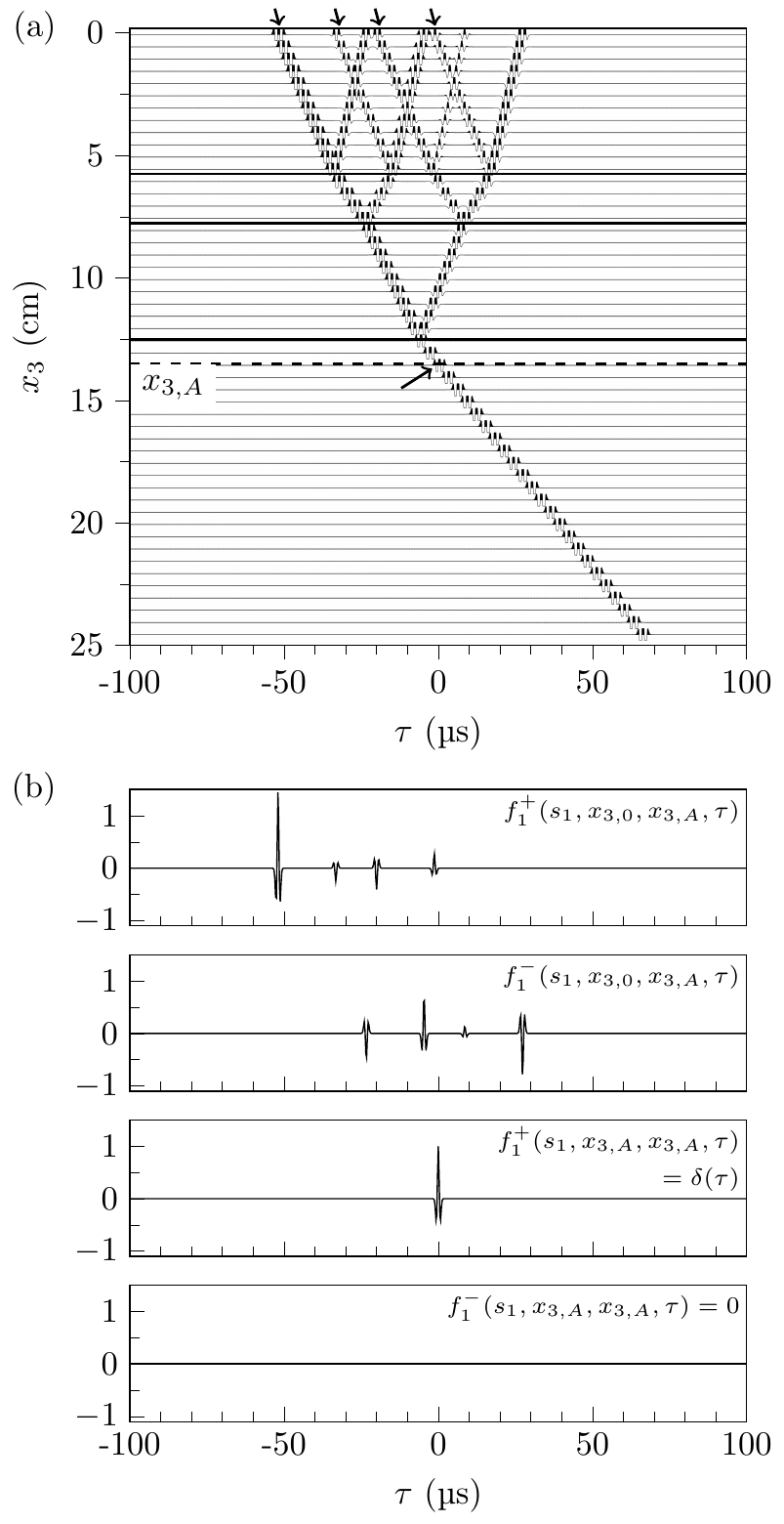}}
\caption{\footnotesize (a) Focusing function $f_1(s_1,x_3,x_{3,A},\tau)$, for $s_1=0.22$ ms/m.  (b) Decomposed focusing functions at $x_{3,0}=0$ and $x_{3,A}$.}
\label{Figure3}
\end{figure}

For horizontally layered media it is convenient to decompose wave fields into plane waves and analyse wave propagation per plane-wave component.
We define the plane-wave decomposition of a wave field quantity $\uu(x_1,x_3,t)$ as
\begin{equation}\label{eqtaup}
\uu(s_1,x_3,\tau)=\int_{-\infty}^\infty \uu(x_1,x_3,\tau+s_1x_1){\rm d}x_1.
\end{equation}
Here $s_1$ is the horizontal slowness and $\tau$ is a new time coordinate, usually called intercept time \cite{Stoffa89Book}. The relation with the more common plane-wave
decomposition by Fourier transform becomes clear if we apply the temporal Fourier transform, $\uu(\omega)=\int_{-\infty}^\infty \uu(\tau)\exp(i\omega\tau){\rm d}\tau$
to both sides of equation (\ref{eqtaup}), which gives
\begin{equation}\label{eqomp}
\u(s_1,x_3,\omega)=\int_{-\infty}^\infty \uu(x_1,x_3,\omega)\exp(-i\omega s_1x_1){\rm d}x_1.
\end{equation}
The tilde denotes the $(s_1,x_3,\omega)$-domain
The right-hand side of equation (\ref{eqomp}) represents a spatial Fourier transform, with wavenumber $k_1=\omega s_1$, where each wavenumber $k_1$ corresponds
to a specific plane-wave component. Similarly, each horizontal slowness $s_1$ in equation (\ref{eqtaup}) refers to a plane-wave component.

Consider an impulsive downgoing plane wave, with horizontal slowness $s_1=0.22$ ms/m, which is incident to the layered medium at $x_{3,0}=0$.
We model its response, employing a $(s_1,x_3,\omega)$-domain modelling method  \cite{Kennett79GJRAS}, adjusted for non-reciprocal media (based
on equations (\ref{eqbb99}) and (\ref{eqbb98}), transformed to the $(s_1,x_3,\omega)$-domain).
The result, transformed back to the $(s_1,x_3,\tau)$-domain,
is shown  in Figure \ref{Figure2}(a) (for fixed $s_1$).
Since it is the response to an impulsive source, we denote this field as a Green's function
$G(s_1,x_3,x_{3,0},\tau)$ (actually Figure \ref{Figure2}(a) shows a band-limited version of the Green's function, in accordance with physical measurements, which are always band-limited).
Note the different angles of the downgoing and upgoing waves directly left and right of the dotted vertical line in the first layer. 
This is a manifestation of the non-reciprocity of the medium. 
Figure  \ref{Figure2}(b) shows the decomposed fields at  $x_{3,0}=0$ and $x_{3,A}$, where  $x_{3,A}$ denotes an arbitrary depth level inside the medium
(taken in this example as $x_{3,A}=13.5$ cm). 
The superscripts $+$ and $-$ stand for downgoing and upgoing, respectively.
For the downgoing field at the upper boundary we have $G^+(s_1,x_{3,0},x_{3,0},\tau)=\delta(\tau)$, where $\delta(\tau)$ is the Dirac delta function. 
For the upgoing response at the upper boundary we write $G^-(s_1,x_{3,0},x_{3,0},\tau)=R(s_1,x_{3,0},\tau)$, where $R(s_1,x_{3,0},\tau)$ is the reflection response. 
This is the response one would obtain from a physical reflection experiment carried out at the upper boundary of the layered medium,
translating it to the moving coordinate system and transforming it to the  plane-wave domain, using equation (\ref{eqtaup}). 
The decomposed responses inside the medium, $G^\pm(s_1,x_{3,A},x_{3,0},\tau)$, which were obtained here by numerical modelling, are not available in a physical experiment. 
In the next section we discuss how these responses can be obtained from  $R(s_1,x_{3,0},\tau)$ using the Marchenko method.
For this purpose, we  introduce an auxiliary wave field, the so-called focusing function $f_1(s_1,x_3,x_{3,A},\tau)$, which is illustrated in Figure \ref{Figure3}(a). 
Here $x_{3,A}$ denotes the focal depth.
The focusing function is defined in a truncated version
of the medium, which is identical to the actual medium above $x_{3,A}$ and homogeneous below $x_{3,A}$.  
The four arrows at the top of Figure \ref{Figure3}(a) indicate the four events of the focusing function leaving the surface $x_{3,0}=0$ as downgoing waves; the arrow just below the dashed line indicates the focus.
Figure  \ref{Figure3}(b) shows the decomposed focusing functions at  $x_{3,0}=0$ and $x_{3,A}$.
The downgoing focusing function $f_1^+(s_1,x_{3,0},x_{3,A},\tau)$ at the upper boundary is designed such that, after propagation through the truncated medium, it focuses at $x_{3,A}$.
The focusing condition at $x_{3,A}$ is $f_1^+(s_1,x_{3,A},x_{3,A},\tau)=\delta(\tau)$.
The upgoing response at the upper boundary is $f_1^-(s_1,x_{3,0},x_{3,A},\tau)$. Because the half-space below
the truncated medium is by definition homogeneous, there is no upgoing response at $x_{3,A}$, hence $f_1^-(s_1,x_{3,A},x_{3,A},\tau)=0$.
\rev{Note that the downgoing and upgoing parts of the focusing function at  $x_{3,0}$ each contain $2^{n-1}$ pulses, where $n$ is the number of interfaces in the truncated medium.}

In a similar way as for reciprocal media \cite{Wapenaar2013PRL, Slob2014GEO}, 
we derive relations between the decomposed Green's functions and focusing functions. 
For this we use general reciprocity theorems for decomposed wave fields $\u^\pm(s_1,x_3,\omega)$ in 
two independent states $A$ and $B$. These theorems read 
\begin{equation}\label{eq142}
\bigl(\u_A^{+\a}\u_B^- - \u_A^{-\a}\u_B^+\bigr)_{x_{3,0}}=
\bigl(\u_A^{+\a}\u_B^- - \u_A^{-\a}\u_B^+\bigr)_{x_{3,A}}
\end{equation}
and
\begin{equation}\label{eq143}
\bigl(\u_A^{+*}\u_B^+ - \u_A^{-*}\u_B^-\bigr)_{x_{3,0}}=
\bigl(\u_A^{+*}\u_B^+ - \u_A^{-*}\u_B^-\bigr)_{x_{3,A}},
\end{equation}
respectively, where superscript $*$ denotes complex conjugation.
These theorems, but without the superscripts $\a$ in equation (\ref{eq142}), were previously derived for reciprocal media \cite{Wapenaar96GJI1}.
Whereas equation (\ref{eq142}) holds for propagating and evanescent waves, equation (\ref{eq143}) only holds for propagating waves.
The extension to non-reciprocal media is derived in the supplementary material.
For non-reciprocal media, the superscript $\a$ at a wave field indicates that this field is defined in the \rev{complementary} medium, in which the coupling parameter $\xi_i$,
appearing in equations (\ref{eqbb99}) and (\ref{eqbb98}),
is replaced by $-\xi_i$. \rev{The terminology ``complementary medium'' is adopted from the literature on non-reciprocal electromagnetic wave theoy \cite{Kong72IEEE, Lindell95JEVA}.}
Note that, when wave fields with a tilde are written without their arguments (as in equations \ref{eq142} and \ref{eq143}), 
it is tacitly assumed that fields indicated by the superscript $\a$ are evaluated at $(-s_1,x_3,\omega)$.

To obtain relations between the decomposed Green's functions and focusing functions,
we now take 
$\u_A^\pm=\tilde f_1^\pm$ and $\u_B^\pm=\tilde G^\pm$.
The conditions at $x_{3,0}$ and $x_{3,A}$ 
discussed above are, in the $(s_1,x_3,\omega)$-domain, 
$\tilde G^+(s_1,x_{3,0},x_{3,0},\omega)=1$, $\tilde G^-(s_1,x_{3,0},x_{3,0},\omega)=\tilde R(s_1,x_{3,0},\omega)$, $\tilde f_1^+(s_1,x_{3,A},x_{3,A},\omega)=1$ and $\tilde f_1^-(s_1,x_{3,A},x_{3,A},\omega)=0$.
Making the appropriate substitutions in equations (\ref{eq142}) and (\ref{eq143}) we thus obtain
\begin{eqnarray}
&&\tilde G^-(s_1,x_{3,A},x_{3,0},\omega)+\tilde f_1^{-\a}(-s_1,x_{3,0},x_{3,A},\omega)\nonumber\\
&&\hspace{.5cm}=\tilde R(s_1,x_{3,0},\omega)\tilde f_1^{+\a}(-s_1,x_{3,0},x_{3,A},\omega)\label{eq240}
\end{eqnarray}
and
\begin{eqnarray}
&&\tilde G^+(s_1,x_{3,A},x_{3,0},\omega)-\{\tilde f_1^+(s_1,x_{3,0},x_{3,A},\omega)\}^*\nonumber\\
&&\hspace{.5cm}=-\tilde R(s_1,x_{3,0},\omega)\{\tilde f_1^-(s_1,x_{3,0},x_{3,A},\omega)\}^*,\label{eq241}
\end{eqnarray}
respectively. 
These representations express the wave field at $x_{3,A}$ inside the non-reciprocal medium in terms of reflection measurements at the surface $x_{3,0}$ of the medium.
These expressions are similar to those in reference \cite{Slob2014GEO}, except that the focusing functions in equation (\ref{eq240}) are defined in the \rev{complementary} medium. Therefore we cannot follow the
same procedure as in  \cite{Slob2014GEO} to retrieve the focusing functions  from equations (\ref{eq240}) and (\ref{eq241}).
To resolve this issue, we derive a symmetry property of the reflection response $\tilde R(s_1,x_{3,0},\omega)$ and use this to obtain a second set of representations.
For the fields at $x_{3,0}$ in states $A$ and $B$ we choose 
$\u_A^+=\u_B^+=1$ and $\u_A^-=\u_B^-=\tilde R$.
Substituting this into the left-hand side of equation (\ref{eq142}) yields $\tilde R(s_1,x_{3,0},\omega)-\tilde R^{\a}(-s_1,x_{3,0},\omega)$. 
We replace $x_{3,A}$ at the right-hand side of equation (\ref{eq142}) by $x_{3,M}$, which is chosen below all inhomogeneities of the medium, so that there are no upgoing waves at $x_{3,M}$.
Hence, the right-hand side of equation (\ref{eq142}) is equal to $0$. We thus find
\begin{eqnarray}
\tilde R^{\a}(-s_1,x_{3,0},\omega)=\tilde R(s_1,x_{3,0},\omega).\label{eq242}
\end{eqnarray}
We obtain a second set of representations by replacing all quantities in equations  (\ref{eq240}) and (\ref{eq241}) by the corresponding quantities in the \rev{complementary} medium. Using equation
 (\ref{eq242}), this yields
 \begin{eqnarray}
&&\tilde G^{-\a}(-s_1,x_{3,A},x_{3,0},\omega)+\tilde f_1^-(s_1,x_{3,0},x_{3,A},\omega)\nonumber\\
&&\hspace{.5cm}=\tilde R(s_1,x_{3,0},\omega)\tilde f_1^+(s_1,x_{3,0},x_{3,A},\omega)\label{eq243}
\end{eqnarray}
and
\begin{eqnarray}
&&\hspace{-.5cm}\tilde G^{+\a}(-s_1,x_{3,A},x_{3,0},\omega)-\{\tilde f_1^{+\a}(-s_1,x_{3,0},x_{3,A},\omega)\}^*\nonumber\\
&&\hspace{.cm}=-\tilde R(s_1,x_{3,0},\omega)\{\tilde f_1^{-\a}(-s_1,x_{3,0},x_{3,A},\omega)\}^*,\label{eq244}
\end{eqnarray}
respectively.

\section{Marchenko method for non-reciprocal media}
In the previous section we obtained four representations, which we regroup into two sets. 
Equations (\ref{eq241}) and (\ref{eq243}) form the first set, containing only focusing functions in the truncated version of the actual medium.
The second set is formed by equations (\ref{eq240}) and (\ref{eq244}), which contain only focusing functions in the truncated version of the \rev{complementary} medium. All equations contain the reflection response 
$\tilde R(s_1,x_{3,0},\omega)$ of the actual medium (i.e., the measured data, transformed to the $(s_1,x_{3,0},\omega)$-domain). 

We now outline the procedure to retrieve the focusing functions and Green's functions from the reflection response, using the Marchenko method. The procedure is similar to that described
in reference \cite{Slob2014GEO}.  For details we refer to this reference; here we emphasize the differences.
The first set of equations, (\ref{eq241}) and (\ref{eq243}), is transformed  from the  $(s_1,x_3,\omega)$-domain to the  $(s_1,x_3,\tau)$-domain.
Using time windows, the Green's functions are suppressed from these equations. Because one of the Green's functions is defined in the actual medium and the other in the \rev{complementary} medium, 
two different time windows are needed, unlike in the Marchenko method for reciprocal media, which requires only one time window. 
Having suppressed the Green's functions, we are left with two equations for the two unknown focusing functions 
$f_1^+(s_1,x_{3,0},x_{3,A},\tau)$ and $f_1^-(s_1,x_{3,0},x_{3,A},\tau)$. These can be resolved from the reflection response $R(s_1,x_{3,0},\tau)$ using the Marchenko method. 
This requires an initial estimate
of the focusing function $f_1^+(s_1,x_{3,0},x_{3,A},\tau)$, which is defined as the inverse of the direct arrival of the transmission response of the truncated medium. 
\rev{In practice we define the initial estimate simply as $\delta(\tau+\tau_{\rm d})$, where $\tau_{\rm d}=\tau_{\rm d} (s_1,x_{3,0},x_{3,A},\tau)$ is the travel time of the direct arrival, which
 can be derived from a background model of the medium. Since we only need a travel time, a smooth background model suffices; no information about the position and strength of the 
 interfaces is needed. }
Once the focusing functions have been found, they can be substituted in the time domain versions of 
equations (\ref{eq241}) and (\ref{eq243}), which yields the Green's functions $G^+(s_1,x_{3,A},x_{3,0},\tau)$ and $G^{-\a}(-s_1,x_{3,A},x_{3,0},\tau)$.
Note that only the retrieved downgoing part of the Green's function, $G^+$,  is defined in the actual medium. Therefore the procedure continues by applying the Marchenko method to the
time domain versions of equations (\ref{eq240}) and (\ref{eq244}). 
This yields the focusing functions $f_1^{+\a}(-s_1,x_{3,0},x_{3,A},\tau)$ and $f_1^{-\a}(-s_1,x_{3,0},x_{3,A},\tau)$ and, subsequently,
the Green's functions $G^{+\a}(-s_1,x_{3,A},x_{3,0},\tau)$ and $G^-(s_1,x_{3,A},x_{3,0},\tau)$. Here the retrieved upgoing part of the Green's function, $G^-$,  is defined in the actual medium.
This completes the procedure for the retrieval of the downgoing and upgoing parts of the Green's functions in the actual medium at depth level $x_{3,A}$ for horizontal slowness $s_1$.
This procedure can be repeated for any slowness corresponding to propagating waves and for any focal depth $x_{3,A}$.

Finally, we discuss how the retrieved Green's functions can be used for imaging.
\rev{Similar as in a reciprocal medium, the relation between these Green's functions in the $(s_1,x_3,\omega)$-domain is 
\begin{equation}
\tilde G^-(s_1,x_{3,A},x_{3,0},\omega)=\tilde R(s_1,x_{3,A},\omega)\tilde G^+(s_1,x_{3,A},x_{3,0},\omega),
\end{equation}
}
where $\tilde R(s_1,x_{3,A},\omega)$ is the plane-wave reflection response at depth level $x_{3,A}$ of the medium below  $x_{3,A}$.
Inverting this equation yields an estimate of the reflection response, according to
\begin{equation}
\langle\tilde R(s_1,x_{3,A},\omega)\rangle=\frac{\tilde G^-(s_1,x_{3,A},x_{3,0},\omega)}{\tilde G^+(s_1,x_{3,A},x_{3,0},\omega)}.\label{eq246}
\end{equation}
Imaging \rev{the reflectivity} at $x_{3,A}$ involves selecting the $\tau=0$ component of the inverse Fourier transform of $\langle\tilde R(s_1,x_{3,A},\omega)\rangle$, hence
\begin{equation}
\langle R(s_1,x_{3,A},\tau=0)\rangle=\frac{1}{2\pi}\int_{-\infty}^\infty\langle\tilde R(s_1,x_{3,A},\omega)\rangle{\rm d}\omega.
\end{equation}
Substituting equation (\ref{eq246}), stabilising the division (and suppressing the arguments of the Green's functions), we obtain
\begin{equation}\label{eqIM}
\langle R(s_1,x_{3,A},0)\rangle=\frac{1}{2\pi}\int_{-\infty}^\infty\frac{\tilde G^-\{\tilde G^+\}^*}{\tilde G^+\{\tilde G^+\}^*+\epsilon}{\rm d}\omega.
\end{equation}

\section{Numerical example}
\begin{figure}
\centerline{\epsfysize=9 cm \epsfbox{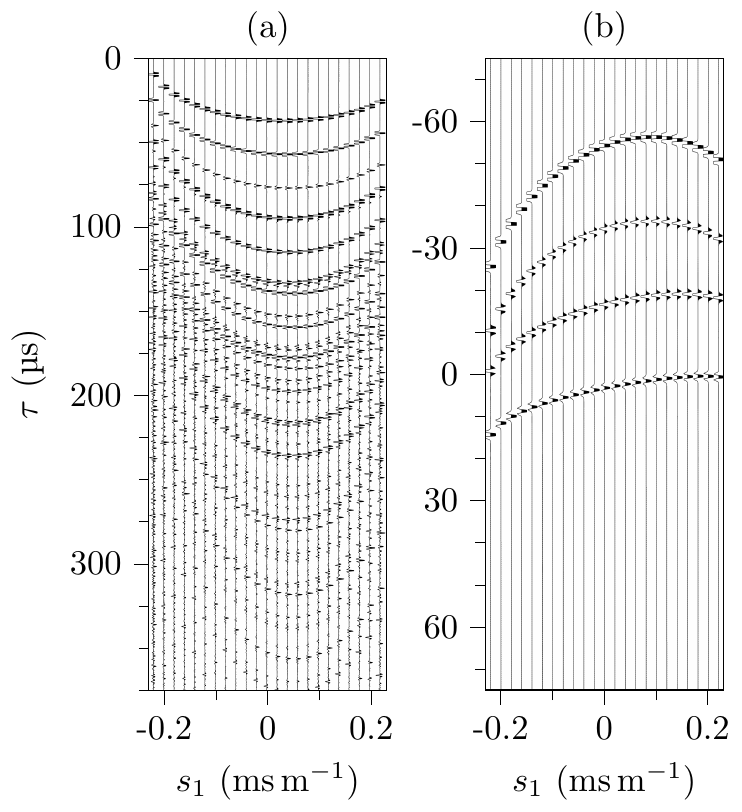}}
\vspace{-0.3cm}
\caption{\footnotesize (a) Modelled reflection response $R(\ssss_1,x_{3,0},\tau)$. (b) Retrieved focusing function $f_1^+(\ssss_1,x_{3,0},x_{3,A},\tau)$. }
\label{Fig4}
\end{figure}

We consider again the layered medium of  Figure \ref{Fig3}.
Using the same modelling approach as before, we model the reflection responses to tilted downgoing plane waves at $x_{3,0}=0$, this time for a range of horizontal slownesses $\ssss_1$. 
The result, transformed to the $(s_1,x_{3,0},\tau)$-domain and convolved with a wavelet with a central frequency of 600 kHz, is shown in Figure \ref{Fig4}(a).
To emphasize the multiples (only for the display), a time-dependent amplitude gain, using the function $\exp\{3\tau/375\mu s\}$, has been applied.
Note the asymmetry with respect to $\ssss_1=0$ as a result of the non-reciprocity of the medium. The last trace (for $s_1=0.22$ ms/m) corresponds with the second trace in Figure \ref{Figure2}(b).

We define the focal depth in the fourth layer, at $x_{3,A}=13.5$ cm. Using the Marchenko method, 
we retrieve the focusing functions $f_1^\pm(\ssss_1,x_{3,0},x_{3,A},\tau)$ and $f_1^{\pm\a}(-\ssss_1,x_{3,0},x_{3,A},\tau)$ from the reflection response $R(\ssss_1,x_{3,0},\tau)$
and the travel times $\tau_{\rm d}$ between $x_{3,0}$ and $x_{3,A}$.
One of these focusing functions, $f_1^+(\ssss_1,x_{3,0},x_{3,A},\tau)$, is shown in Figure \ref{Fig4}(b).
The last trace (for $s_1=0.22$ ms/m) corresponds with the first trace in Figure \ref{Figure3}(b).

\begin{figure}
\centerline{\epsfysize=9cm \epsfbox{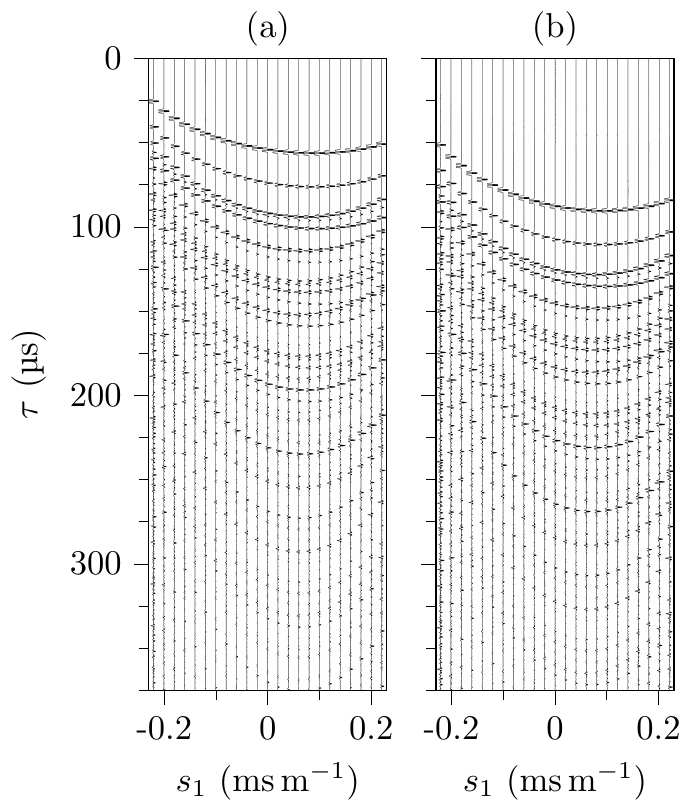}}
\vspace{-0.3cm}
\caption{\footnotesize  (a) Retrieved Green's function $G^+(\ssss_1,x_{3,A},x_{3,0},\tau)$. (b) Idem, $G^-(\ssss_1,x_{3,A},x_{3,0},\tau)$.}
\label{Fig6}
\end{figure}

\begin{figure}
\vspace{-.cm}
\centerline{\epsfysize=10 cm \epsfbox{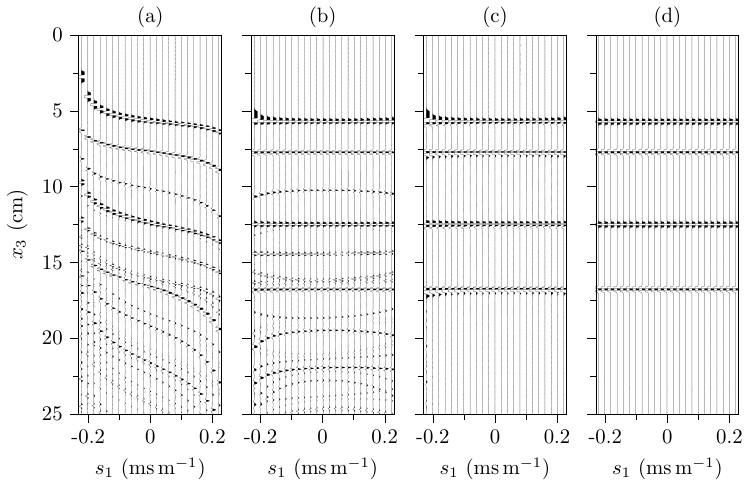}}
\vspace{-0cm}
\caption{\footnotesize  Images of the layered non-reciprocal medium. 
(a) Primary image, accounting for anisotropy but ignoring non-reciprocity. 
(b) Idem, but accounting for non-reciprocity.  (c) Marchenko image. (d) True reflectivity. }
\label{Fig7}
\end{figure}

Using the reflection response and the retrieved focusing functions, 
we obtain the Green's functions $G^+(\ssss_1,x_{3,A},x_{3,0},\tau)$ and $G^-(\ssss_1,x_{3,A},x_{3,0},\tau)$ from the time domain versions of equations 
(\ref{eq241}) and (\ref{eq240}), see Figure \ref{Fig6} (same amplitude gain as in Figure \ref{Fig4}(a)). From the Fourier transform of these Green's functions, an image is obtained
at $x_{3,A}$  as a function of $\ssss_1$, using equation (\ref{eqIM}).
Repeating this for all $x_{3,A}$ we obtain what we call the Marchenko image, shown in Figure \ref{Fig7}(c).
For comparison, Figure \ref{Fig7}(a) shows an image obtained by a primary imaging method, ignoring the non-reciprocal aspects of the medium, and
Figure \ref{Fig7}(b) shows the improvement when non-reciprocity is taken into account (but multiples are still ignored).
For comparison, Figure \ref{Fig7}(d) shows the true reflectivity with the same filters applied as for the imaging  results.
Note that the match of the Marchenko imaging result with the true reflectivity is very accurate. The relative errors, except for the leftmost traces, are less than 2\%.

\rev{Note that we assumed that the medium is lossless. In case of a medium with losses, modifications are required.
For moderate losses that are approximately constant throughout the medium, one can apply a time-dependent loss compensation factor
to the reflection response $R(s_1,x_{3,0},\tau)$ before applying the Marchenko method (assuming an estimate of the loss parameter is available). Alternatively, when the medium is 
accessible from two sides, the Marchenko imaging method of Slob \cite{Slob2016PRL}, modified for non-reciprocal media, can be applied directly to the data.
This removes the need to apply a loss compensation factor.}

\section{Conclusions} 
 We have introduced a new imaging method for layered non-reciprocal materials. The proposed method is a modification of the Marchenko imaging method, 
 which is capable of handling multiple scattering in a data-driven way (i.e., no information is required about the layer interfaces that cause the multiple scattering). 
 To account for  the non-reciprocal properties of the medium, we derived two  sets of representations for the Marchenko method, one set for the actual medium 
 and one set for the \rev{complementary} medium. Using a symmetry relation between the reflection responses of both media, 
 we arrived at a method which retrieves all  quantities needed for imaging (focusing functions and Green's functions in the actual and the \rev{complementary} medium) 
 from the reflection response of the actual medium.
 We illustrated the method with a numerical example, demonstrating the improvement over standard primary imaging methods. 
 The proposed method can be extended for 2D and 3D inhomogeneous media, in a similar way as has been done for the Marchenko method in reciprocal media.

\section{Acknowledgments}
\rev{We thank an anonymous reviewer for a constructive review, which helped us to improve the readability of the paper.}
This work has received funding from the European Union's Horizon 2020 research and innovation programme: European Research Council (grant agreement 742703) and 
Marie Sk\l odowska-Curie (grant agreement 641943).

\newpage
\appendix

\section{An acoustic imaging method for layered non-reciprocal media: Supplementary material}
We derive equations (1), (2), (8) and (9) in the main paper.

\subsection{Acoustic wave equation for a non-reciprocal medium}

The theory for 3D elastic wave propagation in modulated materials, including the homogenisation procedure, is extensively discussed by Nassar et al. \cite{Nassar2017JMPS}.
Here we discuss the main equations, simplified for the acoustic approximation. 
Consider a coordinate system $\bx=(x_1,x_2,x_3)$ that moves along with the modulating wave. 
We start with the following two equations in the space-time $(\bx,t)$ domain 
\begin{eqnarray}
 \partial_t\m_j  &=&-\partial_j\sigmaa ,\label{eq101}\\
\partial_t\Theta&=&\partial_i v_i.\label{eq102}
\end{eqnarray}
Operator $\partial_t$ stands for temporal differentiation and $\partial_i$ for differentiation in the $x_i$-direction. Latin subscripts (except $t$) taken the values 1 to 3.
Einstein's summation convention applies to repeated Latin subscripts, except for subscript $t$.
Equation (\ref{eq101}) formulates equilibrium of momentum in the moving coordinate system (leading order terms only), where 
$\m_j=\m_j(\bx,t)$ is the momentum density and $\sigmaa=\sigmaa(\bx,t)$ the acoustic pressure. 
Equation (\ref{eq102}) relates the  cubic dilatation  $\Theta=\Theta(\bx,t)$ (leading order)
to the particle velocity  $v_i=v_i(\bx,t)$. All field quantities in equations (\ref{eq101}) and (\ref{eq102}) are macroscopic quantities.  
The macroscopic constitutive equations are defined as
\begin{eqnarray}
-\sigmaa&=&K\Theta+S_i^{(1)} v_i,\label{eq103}\\
\m_j&=&S_j^{(2)}\Theta+\rho_{jk} v_k.\label{eq104}
\end{eqnarray}
Here $K=K(\bx)$ is the compression modulus, $\rho_{jk}=\rho_{jk}(\bx)$ the mass density, and $S_i^{(1)}=S_i^{(1)}(\bx)$ and $S_j^{(2)}=S_j^{(2)}(\bx)$ are coupling parameters.
All these coefficients are effective parameters.
Note that the effective mass density is anisotropic, even when it is isotropic at the micro scale. For a lossless non-reciprocal material, the medium parameters are real-valued and
 obey the following symmetry relations
\begin{eqnarray}
\rho_{jk}=\rho_{kj}\quad\mbox{and}\quad S_j^{(2)}=-S_j^{(1)}.
\end{eqnarray}
We rewrite the constitutive equations (\ref{eq103}) and (\ref{eq104}) into explicit expressions for $\Theta$ and $\m_j$, as follows
\begin{eqnarray}
\Theta&=&-\kappa\sigmaa-\xi_i v_i,\label{eq106}\\
\m_j&=&\xi_j\sigmaa+\barrho_{jk} v_k,\label{eq107}
\end{eqnarray}
where
\begin{eqnarray}
\xi_i&=&\kappa S_i^{(1)}\\
\barrho_{jk}&=&\rho_{jk}+\kappa S_j^{(1)}S_k^{(1)},\\
\kappa&=&1/K,
\end{eqnarray}
with $\barrho_{jk}=\barrho_{kj}$. 
Substitution of the modified constitutive equations (\ref{eq106}) and (\ref{eq107}) into equations (\ref{eq102}) and (\ref{eq101}) gives, after some  reorganisation of terms, 
\begin{eqnarray}
 \kappa\partial_t \sigmaa +(\partial_i+\xi_i\partial_t)v_i&=&0,\label{eqbb999}\\
(\partial_j+\xi_j\partial_t) \sigmaa+ {\barrho}_{jk}\partial_tv_k&=&0.\label{eqbb988}
\end{eqnarray}
These are equations (1) and (2) in the main paper.

\subsection{Matrix-vector wave equation}

From here onward we consider a horizontally layered medium, hence, we assume that the medium parameters are functions of the vertical coordinate $x_3$ only, i.e., 
$\kappa=\kappa(x_3)$, $\barrho_{jk}=\barrho_{jk}(x_3)$ and $\xi_i=\xi_i(x_3)$. 
For horizontally layered media it is convenient to decompose wave fields into plane waves and analyse wave propagation per plane-wave component.
We define the Fourier transform from the space-time $(\bx,t)$ domain to the slowness-space-frequency $(\p,x_3,\omega)$ domain as
\begin{eqnarray}
\tilde u(\p,x_3,\omega)=\int\int u(\bx,t)\exp\{\i\omega(t-\p x_\alpha)\}{\rm d}t{\rm d}x_\alpha,\label{eq114}
\end{eqnarray}
where $\p$ denotes the horizontal slowness, $\omega$ the angular frequency and $\i$ the imaginary unit. 
Greek subscripts take on the values 1 and 2 and Einstein's summation convention applies to repeated Greek subscripts. 
Note that equation (\ref{eq114}) accomplishes a decomposition into monochromatic plane waves.

We derive a matrix-vector wave equation of the following form 
\begin{eqnarray}\label{eq22a}
\partial_3\bq =\bA\bq,
\end{eqnarray}
with  wave vector $\bq =\bq (\p,x_3,\omega)$   defined as
\begin{eqnarray}\label{eq23}
\bq =\begin{pmatrix} \tilde\sigmaa \\ \tilde v_3 \end{pmatrix}.
\end{eqnarray}
Equation (\ref{eq22a}) is well-known for wave propagation in reciprocal media\cite{Gilbert66GEO, Frasier70GEO}.
For non-reciprocal media, matrix $\bA$ is obtained as follows. 
From equation (\ref{eqbb999}) we extract an expression for $\partial_3v_3$.
We define $\s_{ij}$ as the inverse of $\barrho_{jk}$, hence, $\s_{ij}\barrho_{jk}=\delta_{ik}$, where $\delta_{ik}$ is the Kronecker delta function.
Applying $\s_{33}^{-1}\s_{3j}$  to  equation (\ref{eqbb988}) yields an expression for $\partial_3\sigmaa$.
By applying $\s_{\alpha j}$ to equation (\ref{eqbb988}) we obtain an expression for $\partial_tv_\alpha$. 
We use equation (\ref{eq114}) to transform these three expressions
to the slowness-frequency domain. In the transformed expressions, $\partial_t$ is replaced by $-\i\omega$ and $\partial_\alpha$  by $\i\omega\p$ for $\alpha=1,2$. 
After elimination of $\tilde v_\alpha$ we thus obtain equation (\ref{eq22a}), with
matrix $\bA =\bA (\p,x_3,\omega)$ defined as
\begin{eqnarray}\label{eq24pe}
\bA=\begin{pmatrix}
 \i\omega\{\xi_3-\ss(\p - \xi_\alpha )\} &\i\omega\s_{33}^{-1} \\
  \i\omega\s_{33}\q^2& \i\omega\{\xi_3-\ss(\p - \xi_\alpha )\}
   \end{pmatrix},
\end{eqnarray}
where
\begin{eqnarray}
\q^2&=&\s_{33}^{-1}\bigl(\kappa-(\ssss_\alpha-\xi_\alpha)\bb_{\alpha\beta}(\ssss_\beta-\xi_\beta)\bigr),\\
\ss&=&\s_{33}^{-1}\s_{3\alpha},\\
\bb_{\alpha\beta}&=&\s_{\alpha\beta}-\s_{\alpha 3}\s_{33}^{-1}\s_{3\beta}.
\end{eqnarray}

\subsection{Decomposition}
We introduce  a decomposed wave vector $\bp =\bp (\p,x_3,\omega)$ via
\begin{eqnarray}
\bq=\bL\bp,\label{eq221}
\end{eqnarray}
where
\begin{eqnarray}\label{eq23a}
\bp =\begin{pmatrix} \u^+ \\ \u^- \end{pmatrix},
\end{eqnarray}
with $\u^+$ and $\u^-$ to be discussed later.
We derive a wave equation for $\bp$, following the same process as for reciprocal media\cite{Kennett79GJRAS, Kennett81GJRAS}, modified for non-reciprocal media. 
The eigenvalue decomposition of matrix $\bA$ reads
\begin{eqnarray}\label{eqALHL}
\bA=\bL\bH\bL^{-1},
\end{eqnarray}
where
\begin{eqnarray}
{\bH}&=&\begin{pmatrix}\i\omega\lambda^+ & 0 \\ 0 & -\i\omega\lambda^-\end{pmatrix},\label{eq122}\\
{\bL}&=&\frac{1}{\sqrt{2}}\begin{pmatrix}1/\sqrt{\h\q} & 1/\sqrt{\h\q}\\  \sqrt{\h\q} &  -\sqrt{\h\q}   \end{pmatrix},\label{eqAL}\\
{\bL}^{-1}&=&\frac{1}{\sqrt{2}}\begin{pmatrix}\sqrt{\h\q} & 1/\sqrt{\h\q}\\  \sqrt{\h\q} &  -1/\sqrt{\h\q}   \end{pmatrix},\label{eq124}
\end{eqnarray}
with
\begin{eqnarray}
\lambda^\pm&=& \q\pm\{\xi_3-\ss(\p - \xi_\alpha)\}, \label{eq127}\\
\q&=&
\sqrt {\s_{33}^{-1}\bigl(\kappa-(\ssss_\alpha-\xi_\alpha)\bb_{\alpha\beta}(\ssss_\beta-\xi_\beta)\bigr)}.\label{eq128}
\end{eqnarray}
%

\begin{figure}
\centerline{\epsfysize=5 cm \epsfbox{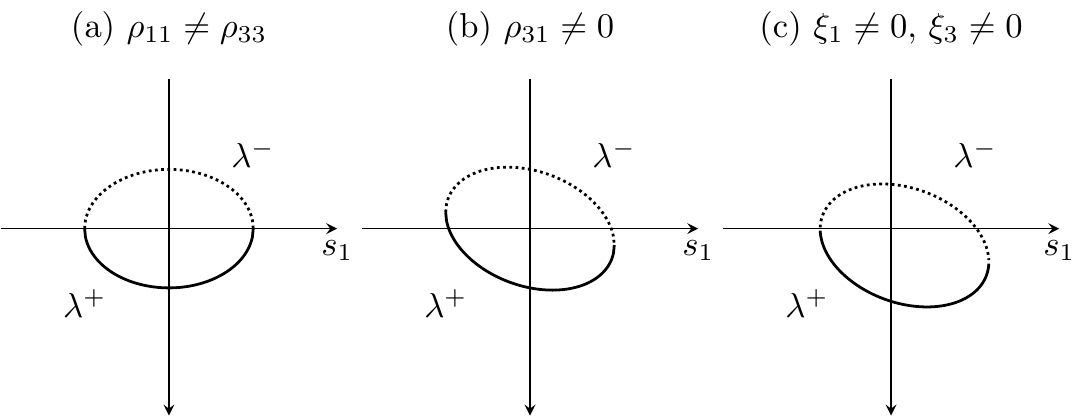}}
\caption{\footnotesize Vertical slowness $\lambda^\pm$ as a function of horizontal slowness $\ssss_1$ (and $\ssss_2=0$).
(a) Anisotropic reciprocal medium. (b) Idem, with tilted symmetry axis. (c) Idem, but for a non-reciprocal medium.}
\label{Fig0}
\end{figure}

Substituting equations (\ref{eq221}) and  (\ref{eqALHL}) into equation (\ref{eq22a}), we obtain
\begin{eqnarray}\label{eq22}
\partial_3\bp =\bB\bp,
\end{eqnarray}
with
\begin{eqnarray}
\bB&=&\bH-\bL^{-1}\partial_3\bL,
\end{eqnarray}
or, using equations (\ref{eq122}) $-$ (\ref{eq124}),
\begin{eqnarray}
\bB&=&\begin{pmatrix}\i\omega\lambda^+ & -r \\ -r & -\i\omega\lambda^-\end{pmatrix},\label{eq131}
\end{eqnarray}
with $\lambda^\pm$ defined in equations (\ref{eq127}) and (\ref{eq128}), and 
\begin{equation}
 r=-\frac{\partial_3(\h\q)}{2\h\q}.\label{eq234}
 \end{equation}
Using equations (\ref{eq23a}) and  (\ref{eq131}), equation (\ref{eq22}) can be written as
\begin{eqnarray}
&&\partial_3\u^+=\i\omega\lambda^+\u^+ - r\u^-,\\
&&\partial_3\u^-=-\i\omega\lambda^-\u^- - r\u^+.
\end{eqnarray}
Analogous to the reciprocal situation, this is a system of coupled one-way wave equations for downgoing waves $\u^+$ and upgoing waves $\u^-$,
with $\lambda^+$ and $\lambda^-$  representing the vertical slownesses for these waves, and $r$ being the reflection function, 
which couples the downgoing waves to the upgoing waves and vice versa.
Figure \ref{Fig0} is an illustration of the vertical slownesses.

\subsection{Propagation invariants}
We consider two independent solutions $\bp_A$ and $\bp_B$ of wave equation (\ref{eq22}) and show that specific combinations of these wave vectors 
 (or ``states'') are invariant for propagation through the medium.
Propagation invariants have been extensively used for wave fields in reciprocal media\cite{Haines88GJI, Kennett90GJI, Koketsu91GJI, Takenaka93WM}. 
To derive propagation invariants for non-reciprocal media, we introduce a complementary medium, in which the coupling parameter $\xi_i$
is replaced by $-\xi_i$ for $i=1,2,3$.
The wave vectors and matrices in a complementary medium are denoted by $\bp^{\a}$ and $\bB^{\a}$, respectively. Using the definition of matrix $\bB$ in equation (\ref{eq131}), with 
$\lambda^\pm$ defined in equations (\ref{eq127}) and (\ref{eq128}) and $r$ in equation (\ref{eq234}), it follows that $\bB$ obeys the following symmetry relations
\begin{eqnarray}
\{\bB^{\a}(-\p,x_3,\omega)\}^t\bN&=&-\bN\bB (\p,x_3,\omega),\label{eq133}\\
\{\bB (\p,x_3,\omega)\} ^\dagger\bJ&=&-\bJ\bB (\p,x_3,\omega),\label{eq134}
\end{eqnarray}
where 
\begin{eqnarray}\label{eq4.3a}
{\bN}=\begin{pmatrix} 0 & 1 \\ -1 & 0 \end{pmatrix},
\quad {\bJ}=\begin{pmatrix} 1 & 0 \\ 0 & -1 \end{pmatrix}.
\end{eqnarray}
Superscript $t$ denotes transposition and $\dagger$ denotes transposition and complex conjugation. Equation (\ref{eq133}) holds for all $\p$, whereas equation
(\ref{eq134}) only holds for those $\p$ for which $\q$ defined in equation (\ref{eq128}) is real-valued, i.e., for $(\ssss_\alpha-\xi_\alpha)\bb_{\alpha\beta}(\ssss_\beta-\xi_\beta)\le \kappa$.
Real-valued $\q$ corresponds to propagating waves, whereas imaginary-valued $\q$ corresponds to evanescent waves.  
We consider the quantities $\partial_3(\{\bp_A^{\a}\}^t\bN\bp_B)$ and $\partial_3(\bp_A^\dagger\bJ\bp_B)$.
When the arguments of functions are dropped, it is tacitly assumed that functions in the complementary medium, indicated by superscript $\a$, are evaluated at $(-\p,x_3,\omega)$.
Applying the product rule for differentiation, using equation (\ref{eq22}) and symmetry
relations  (\ref{eq133}) and  (\ref{eq134}), we find
\begin{eqnarray}
\partial_3(\{\bp_A^{\a}\}^t\bN\bp_B)&=&0\label{eq136}
\end{eqnarray}
and
\begin{eqnarray}
\partial_3(\bp_A^\dagger\bJ\bp_B)&=&0.\label{eq137}
\end{eqnarray}
%
From these equations it follows that $\{\bp_A^{\a}\}^t\bN\bp_B$ and $\bp_A^\dagger\bJ\bp_B$ are independent of $x_3$ (the latter only for propagating waves). These quantities are therefore called propagation invariants.
%
\subsection{Reciprocity theorems}
Using the definitions of $\bp$, $\bN$ and $\bJ$ in equations (\ref{eq23a}) and (\ref{eq4.3a}), equations (\ref{eq136}) and (\ref{eq137}) imply
\begin{equation}\label{eq142b}
\bigl(\u_A^{+\a}\u_B^- - \u_A^{-\a}\u_B^+\bigr)_{x_{3,0}}=
\bigl(\u_A^{+\a}\u_B^- - \u_A^{-\a}\u_B^+\bigr)_{x_{3,A}}
\end{equation}
and
\begin{equation}\label{eq143b}
\bigl(\u_A^{+*}\u_B^+ - \u_A^{-*}\u_B^-\bigr)_{x_{3,0}}=
\bigl(\u_A^{+*}\u_B^+ - \u_A^{-*}\u_B^-\bigr)_{x_{3,A}},
\end{equation}
respectively, where superscript $*$ denotes complex conjugation and $x_{3,0}$ and $x_{3,A}$ denote two depth levels.
These are the reciprocity theorems of equations (8) and (9) in the main paper.


\end{spacing}

\end{document}